\documentclass[UTF-8,preprintnumbers,superscriptaddress,aps,reprint,showkeys]{revtex4}
\usepackage{amsmath,amssymb,graphicx,bm}

\begin{document}

%\color{black}       %% For one column

\title{A new parametrization of Hubble function and Hubble tension}
\author{Tong-Yu He}
\affiliation{College of Physics Science and Technology, Hebei University, Baoding 071002, China}
\author{Jia-Jun Yin}
\affiliation{College of Physics Science and Technology, Hebei University, Baoding 071002, China}
\author{Zhen-Yu Wang}
\affiliation{Yunnan Observatories, Chinese Academy of Sciences, Kunming 650216, China}
\author{Zhan-Wen Han}
\affiliation{College of Physics Science and Technology, Hebei University, Baoding 071002, China}
\affiliation{Yunnan Observatories, Chinese Academy of Sciences, Kunming 650216, China}
\author{Rong-Jia Yang \footnote{Corresponding author}}
\email{yangrongjia@tsinghua.org.cn}
\affiliation{College of Physics Science and Technology, Hebei University, Baoding 071002, China}
\affiliation{Hebei Key Lab of Optic-Electronic Information and Materials, Hebei University, Baoding 071002, China}
\affiliation{National-Local Joint Engineering Laboratory of New Energy Photoelectric Devices, Hebei University, Baoding 071002, China}
\affiliation{Key Laboratory of High-pricision Computation and Application of Quantum Field Theory of Hebei Province, Hebei University, Baoding 071002, China}

%%%%%%%%%%%%%%%%%%%%%%%%%%%%%%%%%%%%  DATE  %%%%%%%%%%%%%%%%%%%%%%%%%%%%%%%%%%%%
%\date{\today}

\begin{abstract}
We present a new parameterized Hubble function and employ observational data from Hubble, Pantheon, and Baryon Acoustic Oscillations to constrain model parameters. The proposed method is thoroughly validated against these datasets, demonstrating a robust fit to the observational data. The obtained best-fit values are $H_0 = 67.5^{+1.3}_{-1.6}$ $\text{km s}^{-1}  \text{Mpc}^{-1}$, $\Omega_{\rm{m0}} = 0.2764\pm{0.0094}$, and $\alpha = 0.33\pm{0.22}$, consistent with the Planck 2018 results, highlighting the existence of Hubble tension.
\end{abstract}
\maketitle

\section{Introduction}
\label{sec1}
Hubble constant ($H_0$) is a fundamental parameter that can quantify the rate of cosmic expansion. It is frequently employed to elucidate the motion of celestial bodies, such as galaxies, relative to the observer's position. In recent decades, measurements of the Hubble constant have garnered attention within the scientific community due to notable discrepancies among results obtained from diverse measurement methodologies and data sources \cite{dacosta2023h0}. Early measurements of the Hubble constant often entailed observations of primordial cosmic signals, such as the cosmic microwave background radiation (CMB). For instance, using the standard cosmological model, $\Lambda$CDM, a Hubble constant of $H_0 = 67.4 \pm 0.5$ $\text{km s}^{-1}  \text{Mpc}^{-1}$ was derived \cite{Planck:2018vyg}. Integrated Baryon Acoustic Oscillations (BAO) measurements with cosmic microwave background (CMB) data from WMAP resulted in $H_{0} = 67.63 \pm 1.30 $ $\text{km s}^{-1}  \text{Mpc}^{-1}$ \cite{Zhang:2018air}. Early Hubble constant measurements consistently indicate values lower than 70 $\text{km s}^{-1}  \text{Mpc}^{-1}$ \cite{Di_Valentino_2021, Anchordoqui_2021}. The age of an old quasar APM 08279+5255 at $z=3.91$ also tends to support a lower Hubble constant \cite{Yang:2009ae}

Later Hubble constant measurements primarily rely on astrophysical observations such as supernovae and galaxies. A result of $H_0 = 73.04 \pm 1.04$ $\text{km s}^{-1} \text{Mpc}^{-1}$ was derived based on the Cepheid-SN (Cepheid-supernova) sample \cite{Riess_2022}. According to \cite{Wong_2019}, they employed a joint analysis of six strongly lensed gravitational lensing events with measured time delays, providing a Hubble constant estimate of $H_0 = 73.3^{+1.7}_{-1.8}$ $\text{km s}^{-1}  \text{Mpc}^{-1}$. The detection of gravitational wave events from neutron star mergers by LIGO and Virgo yielded an estimate of $H_0 = 70^{+12}_{-8}$  $\text{km s}^{-1}  \text{Mpc}^{-1}$ in 2017 \cite{LIGOScientific:2017adf}. Using the Hubble Space Telescope, the Hubble constant was directly measured as $H_0 = 74.03 \pm 1.42 $ $\text{km s}^{-1} \text{Mpc}^{-1}$ through the distance ladder method, providing calibration for the magnitude-redshift relation for 253 Type Ia supernovae \cite{Riess_2011}. It is evident that numerous late-time measurements favor $H_{0} > 70 $ $\text{km s}^{-1} \text{Mpc}^{-1}$ with a minority reporting a value of $H_{0} \approx 70 $ $\text{km s}^{-1} \text{Mpc}^{-1}$.

Observations of the early universe, including CMB data, typically yield lower values for the Hubble constant ($H_0$). Conversely, measurements of celestial bodies at closer distances, such as supernovae and other large-scale cosmic structures, result in higher values for $H_0$. The inconsistency between the results obtained from these two methods has captured the attention of researchers. This disparity is referred to as the Hubble tension \citep{Di_Valentino_2021, DiValentino:2020zio, Wang:2023bxf,Abdalla:2022yfr,Dainotti:2021pqg}, and the statistical significance of these differences surpasses the range of measurement errors, prompting significant discussion and research. The existence of Hubble tension suggests the possibility of unknown physical processes or issues in observational systems concerning the evolution and nature of the universe \citep{Di_Valentino_2021, Verde_2019,Di_Valentino_2019, freedman2017cosmology,Riess_2019}. Efforts to address this issue include improvements in data analysis, reduction of systematic errors, the adoption of new observational methods, and a re-examination of cosmological models.

In addressing the Hubble tension, Lin et al. \cite{Lin_2023} proposed a potential resolution in the Early Dark Sector (EDS), where dark matter mass depends on the Early Dark Energy (EDE) scalar field. They explored a Plank-suppressed EDE coupled with dark matter, finding that this Triggered Early Dark Sector (tEDS) model naturally resolves the coincidence problem of EDE on the background level. Fitting the current cosmological data, including local distance gradients and low-redshift amplitudes of fluctuations, they obtained a Hubble constant of $H_{0} = 71.2$ $\text{km s}^{-1} \text{Mpc}^{-1}$. In the presence of non-standard cosmology, a reconciliation between CMB and local measurements has yielded $H_{0} = 70-74 $ $\text{km s}^{-1} \text{Mpc}^{-1}$ \cite{alcaniz2022hubble}. According to \cite{Yin_2022}, they explored a novel dark fluid model known as the Exponential Acoustic Dark Energy (eADE) model to alleviate the tension in the Hubble telescope. Comparisons with the standard model resulted in $H_0 = 70.06^{+1.13}_{-1.09}$ $\text{km s}^{-1}  \text{Mpc}^{-1}$. Some dynamical dark energy models, see for example \cite{Poulin:2018cxd,Sakstein:2019fmf,Karwal:2021vpk,McDonough:2021pdg,Tutusaus:2023cms,Dahmani:2023goa,Alonso-Lopez:2023hkx,Montani:2023ywn,Torres-Arzayus:2023mdo,
Li:2019yem,Pan:2019gop,Panpanich:2019fxq,DeFelice:2020sdq,Alestas:2020mvb,Alestas:2021luu,Castillo-Santos:2022yoi,Dinda:2021ffa,Sharma:2022oxh,Cai:2021wgv,
Montani:2023xpd,Fang:2024yni,Liu:2023rvo,Liu:2023wew}, could also reduce the Hubble tension.

In this paper, we propose a new Hubble parameterization method and constrain the model parameters using observations from Hubble parameter (Hubble for short later), Pantheon, and BAO data. The results indicate that the Hubble tension may also exist between the Hubble+Pantheon+BAO data and the measurements from local Cepheid–type Ia supernova distance ladder.

The script is structured as follows: In Sec. \ref{s2}, we introduce a new parameterized method for Hubble function. In Sec. \ref{s3}, utilizing the Markov Chain Monte Carlo (MCMC) method, we constrains the cosmological model parameters, namely $H_{0}$, $\Omega_{\rm{m0}}$, and $\alpha$, using the Hubble dataset, the Hubble+Pantheon dataset, and the Hubble+Pantheon+BAO dataset. Sec \ref{s4} presents the result and Sec \ref{s5} is the conclusion of the study.

\section{A new parametrization of Hubble function}
\label{s2}
 According to the Planck 2018 results, the spacetime is spatially flat: $\Omega_{\rm K0}=0.001\pm 0.002$ \cite{Planck:2018vyg}, so here we consider a flat Friedmann-Robertson-Walker-Lema\^{i}tre (FRWL) spacetime
\begin{eqnarray}
\label{frwmet}
ds^2=-dt^2+a^2(t)\left[dr^2+r^2(d\theta^2+\sin^2\theta d\phi^2)\right],
\end{eqnarray}
where $a(t)$ is the scale factor. We use the unit $c=1$ here. For a perfect fluid with energy-momentum tensor, $T_{\mu\nu}=(\rho+p)u_{\mu}u_\nu+pg_{\mu\nu}$, where $\rho$ and $p$ are the energy density and the pressure in the rest frame of the fluid, the 00 component of the Einstein equation, $G_{\mu\nu}=8\pi G T_{\mu\nu}$, yields the Friedmann equation
\begin{eqnarray}
\label{fri}
H^2\equiv \left(\frac{\dot{a}}{a}\right)^2=\frac{8\pi G}{3}\rho_{\rm t},
\end{eqnarray}
where $\rho_{\rm t}$ represents the total energy density contributed from non-relative matter (dark matter and baryonic matter), radiation, and dark energy. Defining dimensionless density parameters $\Omega_{i0}=\rho_{i0}/\rho_{\rm{c}}$, where the subscript `i' denotes a certain energy component, the subscript `0' denotes quantities evaluated today, and the critic density is $\rho_{\rm{c}}=3H^2_0/(8\pi G)$, then the Friedmann equation takes the form
\begin{eqnarray}
\label{F}
&&H^2=H^2_0\left[\Omega_{\rm{m0}}(1+z)^3+\Omega_{\rm{r0}}(1+z)^4+\Omega_{\rm{x}}(z)\right],
\end{eqnarray}
where $\Omega_{\rm{m0}}$ is the nonrelativistic matter (such as dark matter and baryonic matter) density and $\Omega_{\rm{r0}}$ is the radiation density at present time. $\Omega_{\rm{x}}$ represents contributions from dark energy and can be expressed as
\begin{eqnarray}
\label{d}
\Omega_{\rm{x}}(z)=\Omega_{\rm{x0}}\exp\left[3\int_0^z \frac{1+w_{\rm{x}}(z')}{1+z'}dz'\right],
\end{eqnarray}
where $w_{\rm{x}}(z)=p_{\rm{x}}/\rho_{\rm{x}}$ is the equation of state (EoS) of dark energy. For $\Lambda$CDM model, $w_{\rm{x}}(z)=-1$ and $\Omega_{\rm{x}}(z)=1-\Omega_{\rm{m0}}$. For a constant EoS $w_{\rm{x}}(z)\neq -1$, we have
\begin{eqnarray}
\label{w}
&&H^2=H^2_0\left[\Omega_{\rm{m0}}(1+z)^3+\Omega_{\rm{r0}}(1+z)^4+(1-\Omega_{\rm{m0}}-\Omega_{\rm{r0}})(1+z)^{3(1+w_{\rm{x}})}\right].
\end{eqnarray}
This parameterized model is usually called as $w$CDM. When only using low redshift data to limit the model, the contribution of radiation is usually ignored, firstly because the impact of radiation is relatively small, and secondly because if radiation is considered, the fitting results are generally not good.

There are usually two way to parameterize dark energy: one way is to parameterize the EoS (such as the widely used CPL parameterized model \cite{Chevallier:2000qy, Linder:2002dt}), the other is to directly parameterize the Hubble function. Here we adopt the latter approach. Since the Hubble tension mainly rises from Planck 2018 (based on $\Lambda$CDM model) and Cepheid calibrated supernovae Ia measurements \cite{Verde:2019ivm}, we consider a parameterized slightly different from $\Lambda$CDM. Furthermore, since the degeneracy between parameters could affect the fitting results, we consider the model with as few parameters as possible. The Hubble function we suggest takes the following form
\begin{eqnarray}
\label{e}
H^2= H^2_0\left[\Omega_{\rm{m0}}(1+z)^3+(1-\Omega_{\rm{m0}})(1+z)^{\alpha}\right],
\end{eqnarray}
where $\alpha$ is a constant. Hereafter, we denote the parameterized model \eqref{e} as $\alpha$CDM. Comparing Eq. (\ref{e})) and Eq. (\ref{w}), we see that
\begin{eqnarray}
\label{er}
(1-\Omega_{\rm{m0}})(1+z)^{\alpha}=\Omega_{\rm{r0}}(1+z)^4+(1-\Omega_{\rm{m0}}-\Omega_{\rm{r0}})(1+z)^{3(1+w_{\rm{x}})}.
\end{eqnarray}
Comparing with other parameterized models, $\alpha$CDM has only one more parameter than $\Lambda$CDM and unifies the contributions of radiation and dark energy into one term. Due to fewer parameters, $\alpha$CDM could effectively reduce the degeneracies between parameters and so can be used to fit low redshift observational data (because of the limited amount of data, it is difficult to use low redshift observational data to effectively constrain models with more parameters).

In the following data analysis, we will simultaneously constrain the parameters of $w$CDM (where $\Omega_{\rm{r0}}=0$ is assumed) and calculate it's $\alpha$ through the relation $\alpha=3+3 w_{\rm{x}}$ for comparison.

\section{Observational data and methodology}
\label{s3}

In the previous section, we discussed a new Hubble parameterized method, and now we aim to validate whether the approximate values of the model parameters can effectively describe the current universe based on observational data. We primarily employ three datasets, namely the Hubble dataset with 62 data points, the Pantheon dataset with 1701 data points, and a set of six Baryon BAO datasets. For numerical analysis and parameter constraints using the mentioned datasets,

we employ the MCMC method for the global fitting of model parameters. MCMC is a widely-used and effective technique for cosmological parameter estimation. This method allows for thorough exploration of the parameter space, enabling us to find the optimal parameter values while also estimating their uncertainties. The specific steps are as follows: first we define a cosmological model with parameters such as $H_0$, $\Omega_{\rm{m}0}$, and $\alpha$; then by comparing the observed values from each dataset with the theoretical predictions, we construct the likelihood function for each dataset; during the MCMC sampling process, we integrate these datasets by maximizing the likelihood function for Hubble parameter measurements, supernova distance moduli distribution, and BAO peak positions. The use of a combined likelihood function allows us to appropriately weight and combine the error distributions of different datasets, yielding a global likelihood function. Here we utilize the emcee package, which is a Python implementation of MCMC that is especially effective for parallel exploration of multi-dimensional parameter spaces \cite{2010CAMCS...5...65G}. Additionally, to understand the outcomes of the MCMC study, we employ 64 walkers and 2000 steps across all datasets. And the final results will be discussed in the form of 2D contour plots with 1 $\sigma$ and 2 $\sigma$ errors.

\subsection{Observational Hubble data }

Utilizing Hubble observational data to constrain cosmological models is a significant methodology. This approach involves measuring the historical expansion of the universe to derive constraints on cosmological parameters. The Hubble parameter, denoted as $H(z)$, characterizes the rate of cosmic expansion and serves as a fundamental cosmological quantity. Its dependence on redshift ($z$) provides essential insights into the cosmic evolution \cite{2022EPJC...82.1165S,Santos_2016}.
\begin{equation}
H(z)=-\frac{1}{1+z}\frac{dz}{dt}.
\end{equation}
In this part, $dz$ is obtained through spectroscopic surveys, making a measurement of $dt$ a means to derive the model-independent value of the Hubble parameter. The $H(z)$ data set we use consists of 34 $H(z)$ measurements obtained by calculating the differential ages of galaxies, which is called cosmic chronometer \cite{Zhang:2012mp,Stern:2009ep,Moresco:2012jh,Moresco:2016mzx,Ratsimbazafy:2017vga,Stern:2009ep, Moresco:2015cya}, and 28 $H(z)$ measurements inferred
from the BAO peak in the galaxy power spectrum \cite{Gaztanaga:2008xz,Chuang:2012qt,Blake:2012pj,Busca:2012bu,Oka:2013cba,Font-Ribera:2013wce,Anderson:2013oza,Delubac:2014aqe,Wang:2016wjr,Alam:2016hwk,Bautista:2017zgn,Zhao:2018gvb,Borghi:2021rft}, as collected in \cite{Yang:2023qsz}. As the data provided by the DA method is independent of cosmological models, it can be employed to explore alternative cosmological models. The range of these data points is 0 $< z <$ 1.965. Furthermore, we adopt an intermediate value of $H_{0} = 70$ km/s/Mpc for our analysis \cite{2022NewAR..9501659P}. To determine the mean values of the model parameters $H_{0}$, $\Omega_{\rm{m0}}$ and  $\alpha$ (using maximum likelihood analysis), we employ the chi-square function as follows \cite{2012JCAP...03..027G}:
\begin{equation}
\chi _{\rm{H}}^{2}=\sum_{i=1}^{62}\frac{[H_{i}^{\rm{th}}(H_{0},\Omega_{m0},\alpha)-H_{i}^{\rm{obs}}(z_{i})]^{2}}{\sigma _{\rm{H}}^{2}(z_{i})}.
\end{equation}%
The theoretical value of the Hubble parameter is denoted as $H_{i}^{\rm{th}}$, the observed value as $H_{i}^{\rm{obs}}$, and $\sigma _{\rm{H}}^{2}$ represents the standard error of the observed $H\left( z\right) $ values at redshift $z_i$.

\begin{figure}[ht]
\centering
\includegraphics[scale=0.5]{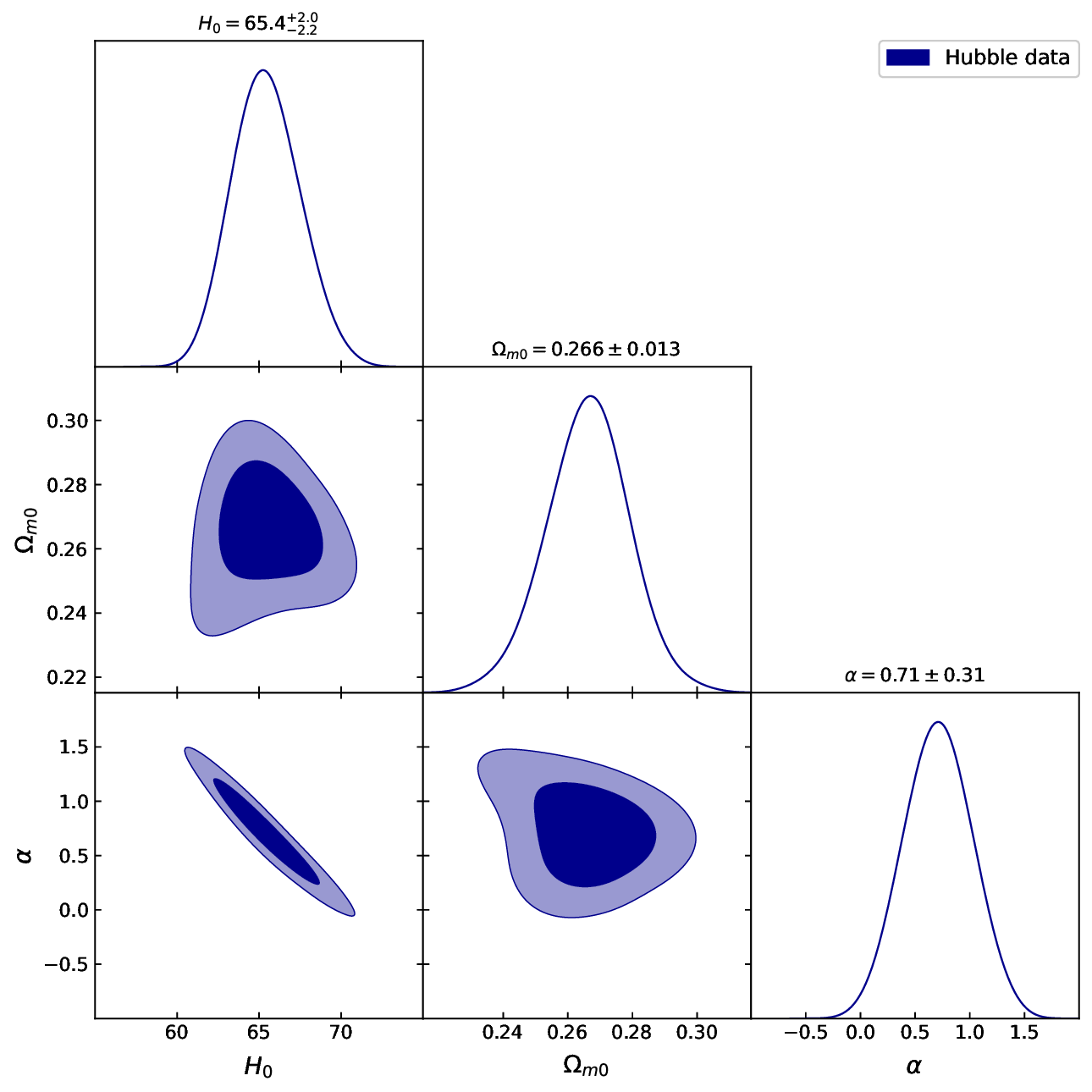}
\caption{The best-fit values for the model parameters with $1 \sigma$ and $2 \sigma$ confidence level contours obtained from Hubble data.}
\label{1}
\end{figure}

As shown in Figure \ref{1}, we obtain the best-fitting values for the model parameters $H_{0}$, $\Omega_{\rm{m0}}$, and $\alpha$, along with 1 $\sigma$ and 2 $\sigma$ confidence level contours. The best-fitting values are  $H_{0}$ = $65.4^{+2.0}_{-2.2}$ $\text{km s}^{-1}  \text{Mpc}^{-1}$, $\Omega_{\rm{m0}}$ = $0.266\pm{0.013}$ and $\alpha$ =$0.71\pm{0.31}$ at 1 $\sigma$ confidence level.

\begin{figure}[ht]
\centering
\includegraphics[scale=0.6]{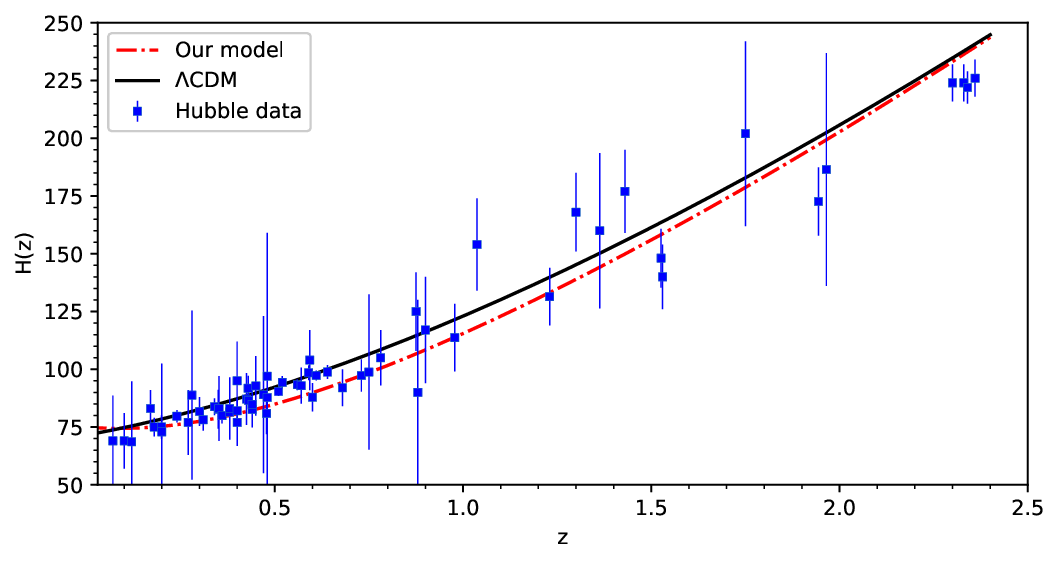}
\caption{The relationship between the Hubble function $H(z)$ and the redshift $z$ obtained from Hubble data. The red line represents $\alpha$CDM and the solid black line corresponds to the $\Lambda$CDM model.}
\label{1.1}
\end{figure}

\begin{widetext}	
\begin{table*}[!htbp]
\begin{center}
%\adjustbox{width=0.5\textwidth}{
\begin{tabular}{l c c c c c c}
\hline\hline
Model & $H_{0}$ $(\text{km/s/Mpc})$ & $\Omega_{\text{m0}}$ & $\alpha$\\
\hline
$\alpha$CDM &  $65.4^{+2.0}_{-2.2}$ & $0.266\pm{0.013}$ &  $ 0.71\pm 0.31$ \\
$w$CDM   & $65.7\pm{2.0}$  & $0.264\pm{0.013}$ & $0.69\pm 0.3$ \\
\hline\hline
\end{tabular}
\caption{The best-fitting values for the model parameters obtained from Hubble data.}
\label{tabh}
\end{center}
\end{table*}
\end{widetext}

In Figure \ref{1.1}, we present error bar plots for the aforementioned Hubble data, compared with the $\Lambda$CDM model ($H_{0} = 67.4$ km/s/Mpc and $\Omega_{\rm{m0}} = 0.315$) \cite{Planck:2018vyg}, indicating that $\alpha$CDM effectively captures the observed Hubble dataset.

We also constrain $w$CDM with the Hubble data and obtain the best-fitting values for the model parameters: $H_{0}=65.7\pm{2.0}$ $\text{km s}^{-1}  \text{Mpc}^{-1}$, $\Omega_{\rm{m0}}=0.264\pm{0.013}$, and $w_{\rm{x}}=-0.77\pm{0.1}$, respectively. Substituting the value of $w_{\rm{x}}$ into the relation: $\alpha=3+3 w_{\rm{x}}$, we get $\alpha=0.69\pm 0.3$ for $w$CDM. We list the best-fitting values in Table \ref{tabh}, and observe that $\alpha$CDM is consistent with $w$CDM within 1 $\sigma$ error, implying that the impact of radiation is relatively weak.

\subsection{Observational Pantheon data}
SNe Ia commonly known as standard candles, serve as powerful distance probes for studying the cosmological dynamics of the universe. Over the past two decades, the sample size of SNe Ia datasets has steadily increased. We employ the largest SNe Ia sample to date,
Pantheon+, which amalgamates data from various surveys such as the Sloan Digital Sky Survey (SDSS), the SNe Legacy Survey (SNLS), the Hubble Space Telescope (HST) survey, and others, comprising 1701 confirmed
supernovae from 18 different surveys \cite{Brout_2022,Riess_2022}. The Pantheon+ dataset spans a redshift range of z $\in$ (0.0012, 2.2614), with a notable increase in the number of SNe at low redshifts. We need to fit the model parameters by comparing the theoretical distance modulus $\mu _{\rm{th}}$ values with the observed $\mu
_{\rm{obs}} $ values. Each distance modulus can be computed using the following formula:
%%%%%%%%%%%%%%%%%%%%%%%%%%%
\begin{equation}
\mu _{\rm{th}}(z)=5\log_{10} \frac{{d_{\rm{L}}(z)}}{\rm{Mpc}}+25 ,
\end{equation}%
where the luminosity distance $d_{\rm{L}}(z)$ is
\begin{equation}
d_{\rm{L}}(z)=(1+z)\int_{0}^{z}\frac{dz^{^{\prime }}}{H(z^{^{\prime }})}.
\end{equation}
The chi-square function $\chi _{\rm{SN}}^{2}$ about the Pantheon data is
\begin{equation}
\chi _{\rm{SN}}^{2}=\sum_{i,j=1}^{1701}\Delta \mu _{i}\left(
C_{\rm{SN}}^{-1}\right) _{ij}\Delta \mu _{j}.  \label{4b}
\end{equation}
In this expression, $C_{\rm{SN}}$ represents
the covariance matrix, as introduced by
Suzuki \cite{Suzuki_2012}. Additionally, $\Delta \mu {i} =\mu
_{\rm{th}}(z_{i},H_{0},\Omega_{\rm{m0}},\alpha )-\mu _{\rm{obs}}$ is defined as the disparity between the observed distance modulus
value, derived from cosmic data, and its theoretical counterpart generated from
the model using the parameter space
$H_{0}$, $\Omega_{\rm{m0}}$ and $\alpha$.

\begin{figure}[ht]
\centering
\includegraphics[scale=0.5]{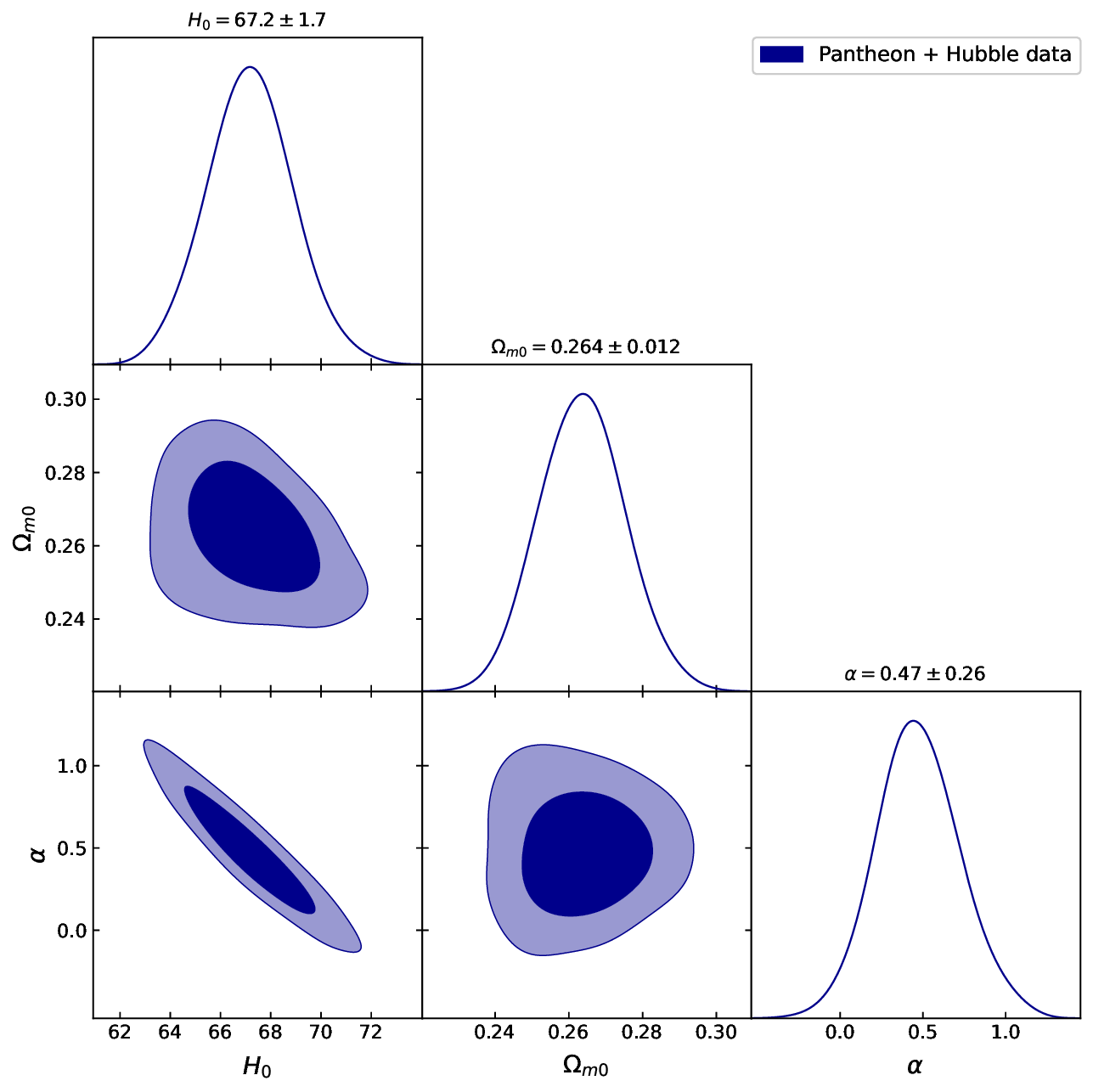}
\caption{The best-fit values for the model parameters with $1 \sigma$ and $2 \sigma$ confidence level contours obtained from Pantheon+Hubble data.}
\label{2}
\end{figure}	

\begin{figure}[ht]
\centering
\includegraphics[scale=0.6]{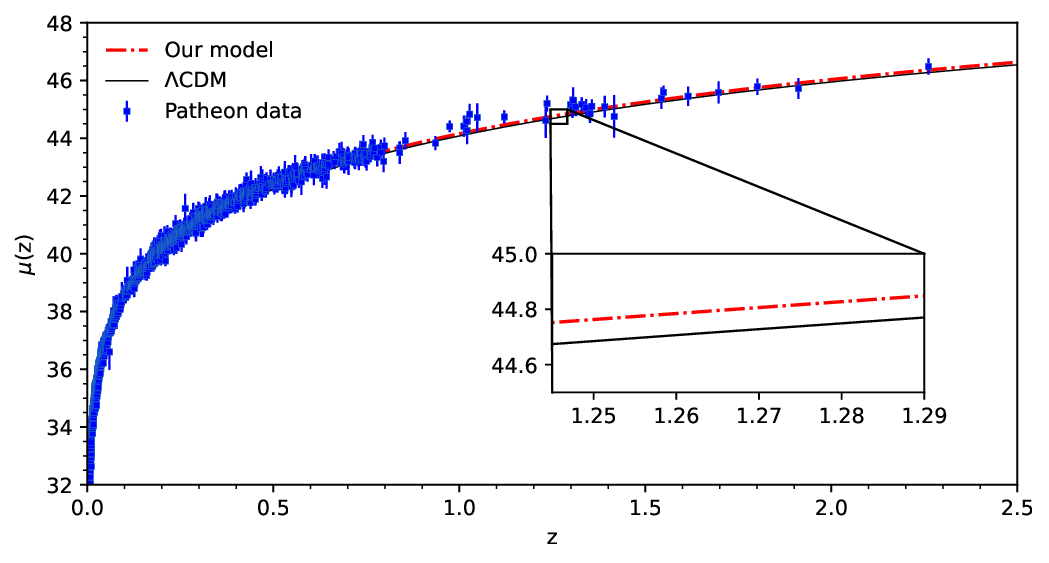}
\caption{The relationship between the distence modulus $\mu(z)$ and the redshift $z$ for $\alpha$CDM and $\Lambda$CDM constrained from Pantheon+Hubble data.}
\label{2.1}
\end{figure}

Taking $\chi_{\rm{H}}^{2} +
\chi_{\rm{SN}}^{2}$ as the minimum constraint for the model parameters $H_{0}$,
$\Omega_{\rm{m0}}$, and $\alpha$, we obtain the best-fit values for these parameters using the Hubble and Pantheon data, as
shown in Figure \ref{2} with 1 $\sigma$ and 2$ \sigma$ confidence level contours. The best-fit
values are $H_{0}$ =
$67.2\pm{1.7}$ $\text{km s}^{-1}  \text{Mpc}^{-1}$, $\Omega_{\rm{m0}}$ =
$0.264\pm{0.012}$ and $\alpha$
=$0.47\pm{0.26}$ at 1 $\sigma$ confidence level.
Additionally, in Figure \ref{2.1}, we
present the error bar plot for the
aforementioned supernova data, comparing $\alpha$CDM with the $\Lambda$CDM model
($H_{0} = 67.4$ km/s/Mpc and $\Omega_{\rm{m0}} = 0.315$) \cite{Planck:2018vyg}. $\alpha$CDM demonstrates a
good fit to the Hubble+Pantheon dataset.

We also use Hubble and Pantheon data to fit $w$CDM, obtaining $H_{0}=67.2\pm 1.8$ $\text{km s}^{-1}  \text{Mpc}^{-1}$, $\Omega_{\rm{m0}}=0.263\pm{0.011}$, and $w_{\rm{x}}=-0.845\pm{0.088}$, respectively. The best-fitting values are listed in Table \ref{tabhpa} for comparison. We observe that $\alpha$CDM is also consistent with $w$CDM within 1 $\sigma$ error, allowed by Hubble+Pantheon data.

\begin{widetext}	
\begin{table*}[!htbp]
\begin{center}
%\adjustbox{width=0.5\textwidth}{
\begin{tabular}{l c c c c c c}
\hline\hline
Model & $H_{0}$ $(\text{km/s/Mpc})$ & $\Omega_{\text{m0}}$ & $\alpha$\\
\hline
$\alpha$CDM &  $67.2\pm{1.7}$ & $0.264\pm{0.012}$ &  $0.47\pm{0.26}$ \\
$w$CDM   & $67.2\pm 1.8$ & $0.263\pm{0.011}$  & $0.465\pm{0.264}$ \\
\hline\hline
\end{tabular}
%}
\caption{The best-fitting values for the model parameters obtained from Hubble+Pantheon data.}
\label{tabhpa}
\end{center}
\end{table*}
\end{widetext}

\subsection{Observational Baryon Acoustic Oscillations data}

BAO arises from acoustic density waves in the early universe's primordial plasma, causing fluctuations in the observable baryonic matter density in the cosmos.
Detecting BAO involves large-scale surveys and redshift measurements to gather information about the large-scale structure of the universe. BAO detectors offer highly precise measurements of large-scale structures, largely unaffected by uncertainties in the nonlinear evolution of the matter density field and other systematic errors. They are considered a standard ruler for measuring the cosmological background evolution \cite{2018APJ}. To improve statistical significance, broaden the redshift range, and gain more comprehensive cosmological insights, we utilize the combined data from six different BAO measurements at various redshifts \cite{Beutler_2011,Alam_2017,Myrzakulov_2023}.
The information taken from the BAO peaks in the matter power spectrum can be used to determine the Hubble parameter $H(z)$ and the angular diameter distance $d_{A}(z)$ which takes the form \newline
\begin{equation}
d_{A}(z)=\int_{0}^{z}\frac{dz^{^{\prime }}}{H(z^{^{\prime }})}.  \label{4d}
\end{equation}%
The combination of the angular diameter distance and the Hubble parameter, $D_{V}(z)$, is determined by \cite{SDSS:2005xqv}
\begin{equation}
D_{V}(z)=\left[ d_{A}(z)^{2}z/H(z)\right] ^{1/3}.
\end{equation}
The chi-square function $\chi _{\rm{BAO}}^{2}$ about BAO is given by the following expression
\begin{equation}
\chi _{\rm{BAO}}^{2}=X^{T}C_{\rm{BAO}}^{-1}X,
\end{equation}%
where
\begin{widetext}
\begin{equation*}
X=\left(
\begin{array}{c}
\frac{d_{A}(z_{\star })}{D_{V}(0.106)}-30.95 \\
\frac{d_{A}(z_{\star })}{D_{V}(0.2)}-17.55 \\
\frac{d_{A}(z_{\star })}{D_{V}(0.35)}-10.11 \\
\frac{d_{A}(z_{\star })}{D_{V}(0.44)}-8.44 \\
\frac{d_{A}(z_{\star })}{D_{V}(0.6)}-6.69 \\
\frac{d_{A}(z_{\star })}{D_{V}(0.73)}-5.45%
\end{array}%
\right).
\end{equation*}%
The inverse covariance matrix $C_{\rm{BAO}}^{-1}$ is
represented in \cite{Myrzakulov_2023}. The six BAO datasets are provided in Table \ref{tab1}. At $z_{\ast } \simeq 1091$, photon decoupling occurred, allowing the CMB to propagate through the universe, eventually becoming the observed cosmic microwave background radiation today. This redshift value is derived through detailed observations and analysis of the CMB \cite{Solanki_2021}.
\begin{table}[]
\begin{center}
\begin{tabular}{ccccccc}
\hline\hline
$z_{\rm{BAO}}$ & $0.106$ & $0.2$ & $0.35$ & $0.44$ & $0.6$ & $0.73$ \\ \hline
$\frac{d_{A}(z_{\ast })}{D_{V}(z_{\rm{BAO}})}$ & $30.95\pm 1.46$ & $17.55\pm 0.60$
& $10.11\pm 0.37$ & $8.44\pm 0.67$ & $6.69\pm 0.33$ & $5.45\pm 0.31$ \\
\hline\hline
\end{tabular}
\caption{Values of $d_{A}(z_{\ast })/D_{V}(z_{\rm{BAO}})$ for distinct values of $z_{\rm{BAO}}$.}
\label{tab1}
\end{center}
\end{table}
\end{widetext}

\begin{figure}[ht]
\centering
\includegraphics[scale=0.5]{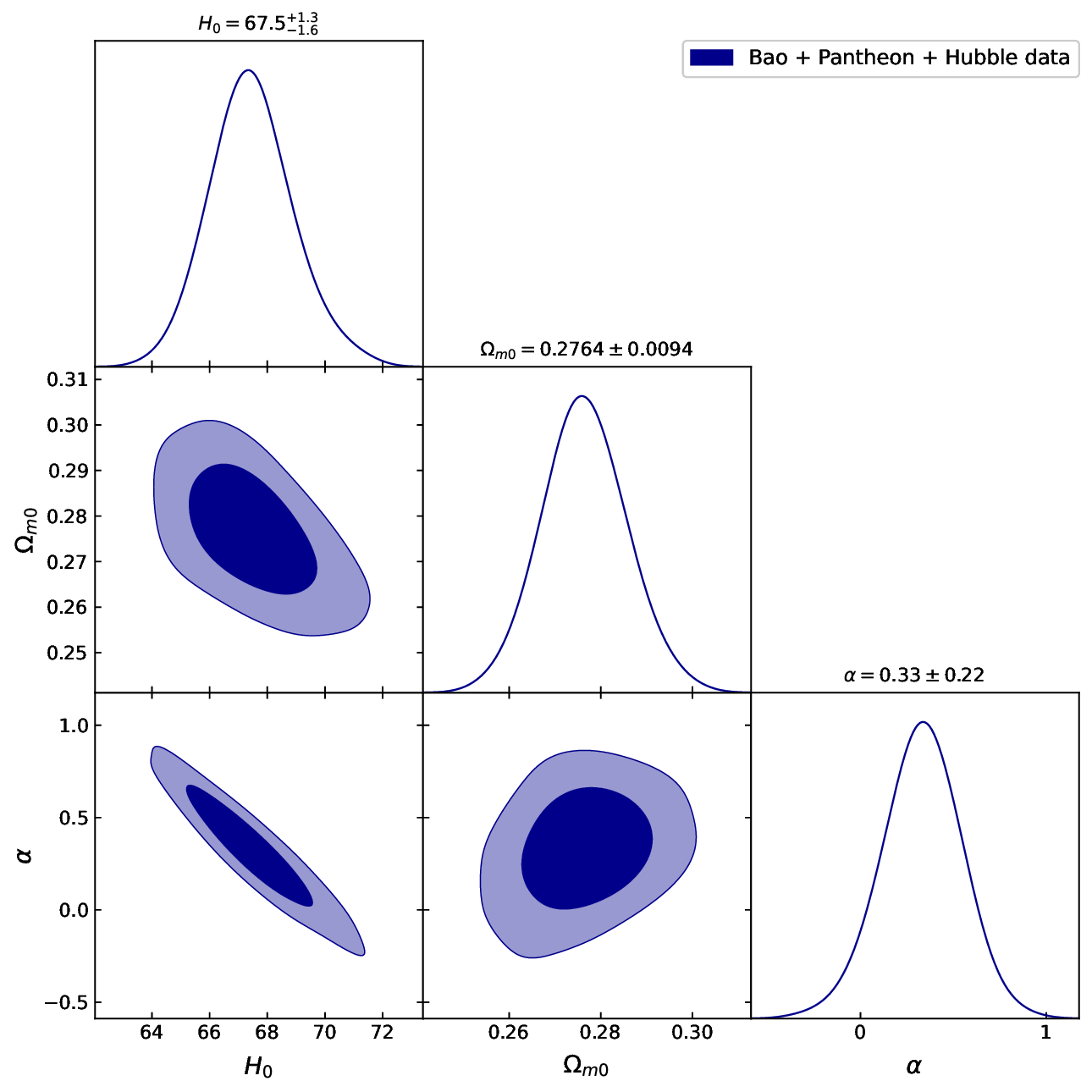}
\caption{The best-fit values for the model parameters with $1 \sigma$ and $2 \sigma$ confidence level contours obtained from the Hubble+Pantheon+BAO data.}
\label{3}
\end{figure}

By minimizing $\chi_{\text{H}}^{2} + \chi_{\text{SN}}^{2}$ + $\chi _{\text{BAO}}^{2}$ to constrain the model parameters $H_{0}$, $\Omega_{\text{m0}}$, and $\alpha$, we determine the best-fit values using the Hubble+Pantheon+BAO dataset, as illustrated in Figure \ref{3}. The resulting values are $H_0 = 67.5^{+1.3}_{-1.6}$ $\text{km s}^{-1}  \text{Mpc}^{-1}$, $\Omega_{\rm{m0}} = 0.27684\pm{0.0094}$, and $\alpha = 0.33\pm{0.22}$, at 1 $\sigma$ confidence level.

\begin{widetext}	
\begin{table*}[!htbp]
\begin{center}
%\adjustbox{width=0.5\textwidth}{
\begin{tabular}{l c c c c c c}
\hline\hline
Model & $H_{0}$ $(\text{km/s/Mpc})$ & $\Omega_{\text{m0}}$ & $\alpha$\\
\hline
$\alpha$CDM &  $67.5^{+1.3}_{-1.6}$ & $0.2764\pm{0.0094}$ &  $ 0.33\pm{0.22}$ \\
$w$CDM  & $67.3\pm{1.4}$ & $0.2772\pm{0.0098}$ & $0.345\pm{0.22}$ \\
\hline\hline
\end{tabular}
%}
\caption{The best-fitting values for the model parameters obtained from Hubble+Pantheon+BAO data.}
\label{tabhp3}
\end{center}
\end{table*}
\end{widetext}

We also use Hubble+Pantheon+BAO data to constrain $w$CDM, obtaining $H_{0}=67.3\pm{1.4}$ $\text{km s}^{-1}  \text{Mpc}^{-1}$, $\Omega_{\rm{m0}}=0.2772\pm{0.0098}$, and $w_{\rm{x}}=-0.885\pm 0.072$, respectively. For comparison, we list the best-fitting values in Table \ref{tabhp3} and observe that $\alpha$CDM is still consistent with $w$CDM within 1 $\sigma$ error, again implying that the impact of radiation is relatively weak, allowed by Hubble+Pantheon+BAO data.

\section{Results}
\label{s4}

In this section, We discuss the results in Sec \ref{s3}. We leveraged multiple observational datasets, including Hubble, Hubble+Pantheon samples and Hubble+Pantheon+BAO data \cite{Giostri_2012}. Employing the MCMC method, we finally constrained the model parameters $H_0$, $\Omega_{\rm{m0}}$, and $\alpha$.

Firstly, the Hubble dataset, consisting of 62 data points, was utilized to quantify the historical expansion of the universe. In Figure \ref{1}, the best-fitting values for the model parameters were determined as $H_{0} =65.4^{+2.0}_{-2.2}$ $\text{km s}^{-1}  \text{Mpc}^{-1}$, $\Omega_{\rm{m0}} = 0.266\pm{0.013}$ , and $\alpha = 0.71\pm{0.31}$. It can be observed that the fitting parameters $H_{0}$ have relatively large uncertainties, and a significant contributing factor is the limited amount of data. In Figure \ref{1.1}, we can find that our model is relatively close to the $\Lambda$CDM model, and it fits well with the observational data. Next, in order to reduce the uncertainties in the parameters, we have incorporated 1701 observational data points from the Pantheon dataset \cite{2022NewAR..9501659P}. Figure \ref{2} presents the fitting results: $H_{0} =67.2\pm{1.7}$ $\text{km s}^{-1}  \text{Mpc}^{-1}$, $\Omega_{\rm{m0}}$ = $0.264\pm{0.012}$ and $\alpha$ =$0.47\pm{0.26}$. It is observed that the error in $H_0$ has decreased. In Figure \ref{2.1}, $\alpha$CDM is seen to fit well with the observational data from Pantheon and is close to the $\Lambda$CDM model. Next, we included an additional 6 BAO data points and obtained the fitting results: $H_0 = 67.5^{+1.3}_{-1.6}$ $\text{km s}^{-1}  \text{Mpc}^{-1}$, $\Omega_{\rm{m0}} = 0.2764\pm{0.0094}$, and $\alpha = 0.33\pm{0.22}$. In Table \ref{tab2}, we can find with the increase in the observational data size, the constraints on the parameters $H_{0}$, $\Omega_{\rm{m0}}$, and $\alpha$ become progressively more accurate. To validate the model's efficacy, we compared error bar plots for Hubble and Pantheon datasets with the $\Lambda$CDM model. Consistently, the model fits the observed dataset very well.

\begin{widetext}	
\begin{table*}[!htbp]
\begin{center}
%\adjustbox{width=0.5\textwidth}{
\begin{tabular}{l c c c c c c}
\hline\hline
Data   & $H_{0}$ $(\text{km/s/Mpc})$ & $\Omega_{\text{m0}}$ & $\alpha$ \\
\hline
Prior   & $(64.9, 76.8)$  \\
Hubble & $65.4^{+2.0}_{-2.2}$ & $0.266\pm{0.013}$ &  $0.71\pm{0.31}$ \\
Hubble+Pantheon   & $67.2\pm{1.7}$ & $0.264\pm{0.012}$  & $0.47\pm{0.26}$ \\
Hubble+Pantheon+BAO   & $67.5^{+1.3}_{-1.6}$ & $0.2764\pm{0.0094}$ & $0.33\pm{0.22}$ \\
\hline\hline
\end{tabular}
%}
\caption{A summary of the results derived from the analysis of three datasets.}
\label{tab2}
\end{center}
\end{table*}
\end{widetext}

From Table V, we observe that the fitted values of $H_0$ using different datasets are slight variations, but are relatively close, indicating that $H_0$ is not highly sensitive to the choice of dataset. The fitted values of $\Omega_{\rm{m}0}$ show minimal variation, ranging from 0.264 to 0.2764, suggesting low sensitivity to dataset choice. However, the fitted values of $\alpha$ vary more significantly across different datasets, ranging from 0.71 to 0.33, indicating higher sensitivity and greater uncertainty for this parameter. Compared to $\Lambda$CDM constrained from Pantheon+SH0ES \cite{Brout_2022} and the model proposed in \cite{Li:2019yem} with two parameters constrained from Pantheon+BAO data, the values of $H_0$ differ significantly from the values obtained here, meaning the results are sensitive to the choice of model parameters. Compared to $w$CDM constrained here and the model proposed in \cite{Myrzakulov_2023} with three parameters constrained from Hubble+Pantheon+BAO data, the values of $H_0$ differ slightly from the values obtained here, implying the results may be not sensitive to the choice of model parameters.

\section{Conclusions}
\label{s5}

In conclusion, we proposed a new Hubble parameterization method and constrained the model parameters using observations from Hubble, Pantheon, and BAO. This model is validated with Hubble, Pantheon, and BAO data, providing best-fit values for $H_0 = 67.5^{+1.3}_{-1.6}$ $\text{km s}^{-1}  \text{Mpc}^{-1}$, $\Omega_{\rm{m0}} = 0.2764\pm{0.0094}$, and $\alpha = 0.33\pm{0.22}$, consistent well with the Planck 2018 data ($H_{0} = 67.4\pm 0.5$ Km/s/Mpc \cite{Planck:2018vyg}), but deviating from Cepheid-supernova observation ($H_{0} = 73.04\pm 1.04$ $\text{km s}^{-1}  \text{Mpc}^{-1}$ \cite{Riess_2022}) by more than 4 $\sigma$. The Hubble tension was mainly triggered with the higher Hubble constant ($H_0=74.0\pm 1.04$ $\text{km s}^{-1} \text{Mpc}^{-1}$ \cite{Riess_2022}) estimated from the local Cepheid–type SNe Ia distance ladder being at odds with the lower value extrapolated from CMB data, assuming the standard $\Lambda$CDM cosmological model ($H_0=67.4\pm 0.5$ $\text{km s}^{-1}  \text{Mpc}^{-1}$ \cite{Planck:2018vyg}). However, our analyses indicate that the Hubble tension may also exist between the Hubble+Pantheon+BAO data and the measurements from local Cepheid–type Ia supernova distance ladder if taking the parametrization \eqref{e}. We also constrained $w$CDM with the same datasets for comparison and found that $\alpha$CDM is consistent with $w$CDM within 1 $\sigma$ error, indicating that the impact of radiation is relatively weak.

Any potential biases or systematic errors in the observational data may have some impact on the results here. Completely eliminating biases and systematic errors in observational data is challenging and impossible. The datasets we used have undergone extensive cross-calibration and systematic error corrections by other researchers to minimize residual biases and observational errors \cite{Brout_2022,Riess_2022}. Additionally, we accounted for these residual errors in our simulations and analyses by incorporating error terms in the Monte Carlo likelihood function and performing error propagation analysis. Using multiple datasets also aids in cross-verifying results, thereby reducing the influence of systematic errors from any single dataset. Table \ref{tab2} provides the upper and lower limits for each parameter, demonstrating the sensitivity of our results to these errors. These bounds indicate that the impact of these errors is limited and does not significantly affect the overall conclusions of our analysis.

These results may contribute to our understanding of the current universe and its cosmodynamics. Future research directions may encompass the expansion of datasets, exploration of additional cosmological models, and investigation into the impact of observational technologies on parameter constraints. The continuous advancement in observational methods is anticipated to refine our understanding of the universe's evolution.

\section*{Acknowledgments}
This study is supported in part by National Natural Science Foundation of China (Grant No. 12333008) and Hebei Provincial Natural Science Foundation of China (Grant No. A2021201034).

\bibliographystyle{apsrev}%{elsarticle-num}
\bibliography{a}
%\end{thebibliography}

\end{document}